\documentclass[11pt,twoside]{article}
\usepackage{asp2010}

\resetcounters
\begin{document}

\title{Query Driven Visualization}
\author{Hugo Buddelmeijer$^1$ and Edwin A. Valentijn$^1$
\affil{$^1$Kapteyn Astronomical Institute, Postbus 800, 9700 AV Groningen, The Netherlands}
}

\begin{abstract}
The request driven way of deriving data in Astro-WISE is extended to a query driven way of visualization. 
This allows scientists to focus on the science they want to perform, because all administration of their data is automated.
This can be done over an abstraction layer that enhances control and flexibility for the scientist.
\end{abstract}

\section{Query Driven Visualization}
Ultimately, astronomers answer questions about the nature of the universe.
Visualization and exploration of data is an essential tool in this process.
Traditionally this analysis phase is based on the data that is made available through an earlier process.
In this paper we turn this around with \textit{query driven visualization}: the data processing is based on what is necessary for the requested visualization.

With query driven visualization there is a close interaction between the visualization software and the software responsible for storing and processing the data.
We use the term \textit{information system} to refer to the combination of all software dealing with the data, even though no formal connection between the individual components is necessary.
In particular we refer to {\sf Astro-WISE} \citep{P160_adassxxi}, although the presented research is applicable to other information systems as well.

The presented methodology allows scientists to request data directly with their visualization software, either explicitly or through interaction.
The information system will subsequently provide the data required for the visualization automatically in an optimal way.
This allows the visualization software to focus on displaying the data, and the scientist on the questions he or she wants to answer.

\section{Target Processing}
The basis of our query driven visualization methods is the request driven way of processing developed for 
{\sf Astro-WISE}, called \textit{target processing}.
A \textit{Target} in {\sf Astro-WISE} is a representation of a science product and can be seen as an object in Object Oriented programming sense.
Accessing a specific science product, e.g. a catalog, amounts to requesting the Target that represents the science product.
The information system will autonomously determine whether there is a suitable existing Target that fulfills the request directly, or what is required to derive it otherwise.

Targets not only represents specific data sets, but also the process to create this data.
Every Target is of a specific class that describes what kind of science product the Target represents.
This class forms a blueprint for the creation of such a science product and prescribes how to derive the data from other Targets and what parameters can be set to influence this processing.
The Targets themselves are stored with all the details required to process them at any time for any reason.
In particular, links to all other Targets that are used as input are stored, these are called \textit{dependencies}.

It is important for this paper to make the distinction between the creation and use of a Target itself and the creation and use of the data it represents.
A Target is considered to exist as soon as its dependencies and process parameters are set, and can be used and stored from thereon.
A Target can be processed partially and the processing result might be stored locally, at a remote dataserver or not at all.
The processing result might therefore not be available.
This allows Targets to be created as general as possible to maximize their reusability while retaining scalability.

\section{Dependency Graphs}
The dependencies of a Target are other Targets from which it is derived, which will have dependencies of their own \citep{Mwebaze:2009:ATU:1683300.1683752}.
The set of dependencies that links a Target all the way back to the raw data is called a \textit{dependency graph} (or \textit{tree}).
The information system will create a dependency graph autonomously to fulfill a request for data.

The information system will discover and reuse existing Targets as much as possible.
These existing Targets could have been created by other scientist, resulting in implicit sharing of data.
New Targets are created to be as reusable as possible for future requests.
This means that newly created Targets might represent more data than is strictly necessary to fulfill this specific request.
The Targets in this dependency graph are subsequently stored, thereby storing the graph itself, without being processed.

Only parts of the Targets in the dependency graph are necessary to fulfill the request, that is, to create the data of the end node.
A novel way to process Targets partially is implemented in {\sf Astro-WISE} in order to prevent the creation of unnecessary catalog data.
This partial processing is done in an implicit way through optimization of the dependency graph.
The information system will modify the dependency graph by temporarily substituting parts of it with different Targets.
This is done such that the modified tree can be processed more efficiently than the original, while ensuring that the final Target still represents the requested science product.
The resulting dependency graph contains Targets that represent subsets of the Targets in the original graph.
These transient Targets are processed in full, and the resulting data is stored as part of the original Targets, but only if beneficial for performance.
This results in the required scalability without concern for the scientist or visualization software.

\begin{figure}[ht!]
 \centering
 \includegraphics[width=0.8\linewidth]{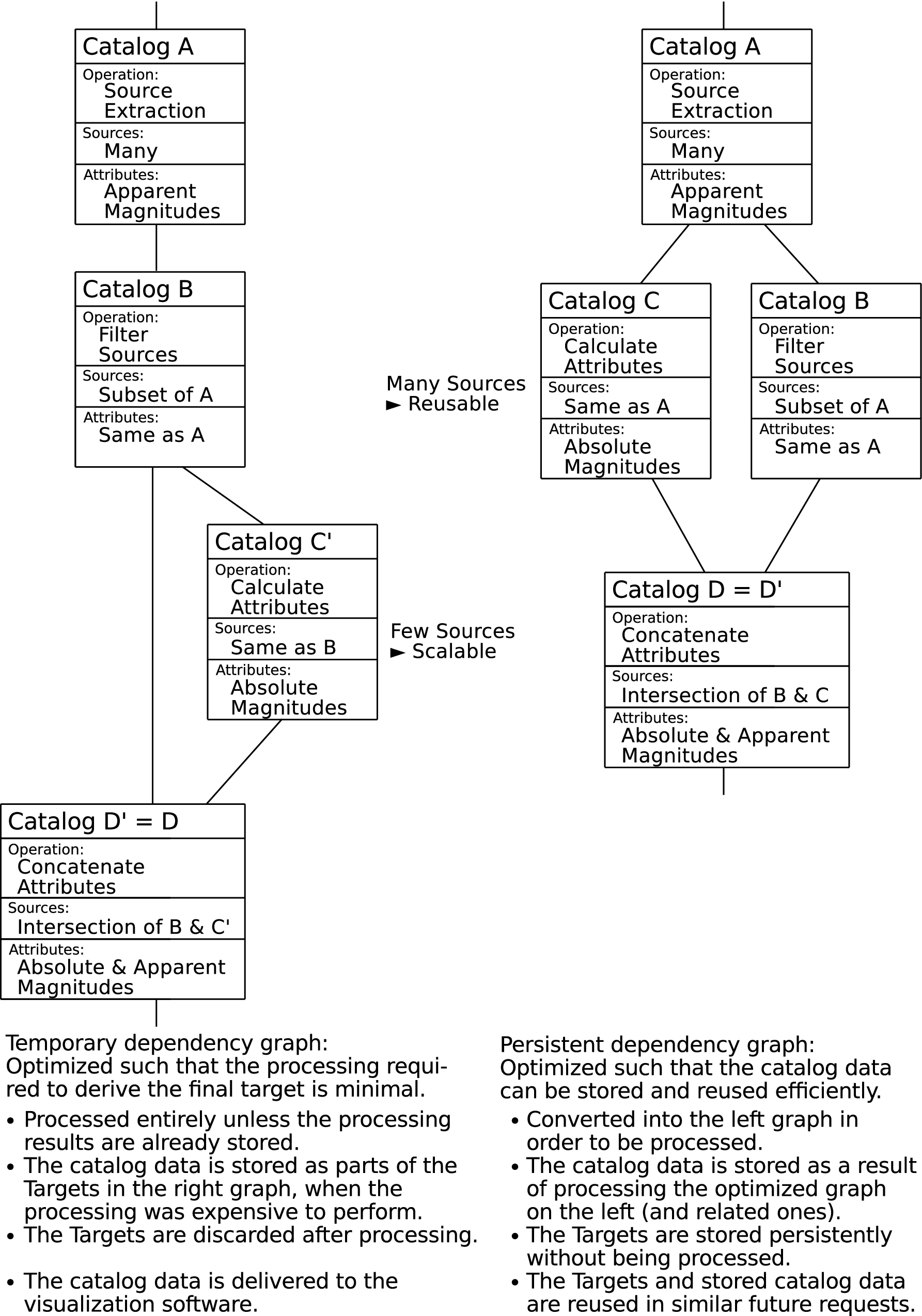}
 \caption{Two dependency graphs as generated automatically by the information system when attributes for a subset of the sources of catalog $A$ are requested.
Every box is a Target that represents a catalog, derived from the catalog above it.
The two dependency graphs are equivalent: their final Targets ($D'$ and $D$) represent the same catalog. The graph on the left is used to create the catalog data in the most efficient way, in this case by processing Target $B$ on the database and $C'$  and $D'$ on the scientist's workstation.
The graph on the right is stored persistently and is used to store catalog data such that it can easily be reused, in this case by storing the catalog data of $C'$ in the database as part of $C$. For example, catalog $C$ (and thus the processing result of $C'$) can be reused when absolute magnitudes are requested for a different subset of $A$. As a result, catalogs are created with maximal reuse while processed with maximal scalability, without requiring intervention by the scientist or visualization software.}
 \label{fig:scintroexample2}
\end{figure}

An example of a dependency graph in both forms (optimized for reuse and sharing, and optimized for processing and scalability) is given in figure \ref{fig:scintroexample2}.
This example is explained more thoroughly in \citet{pullcatalogpaper}.

\section{Abstraction}
Target processing is well suited for abstraction, due to its declarative way of requesting and processing data.
Furthermore, all the details of the processing are specified in a standardized form.

The Simple Application Messaging Protocol (SAMP) is a standard of the International Virtual Observatory Alliance (IVOA) for interaction between astronomical applications through application-defined messages.
New messages are proposed to perform query driven visualization and target processing in general over SAMP \citep{qdvpaper}.

This interaction can be performed on various levels, depending on the level of knowledge the visualization software has of the underlying information system.
Some applications only need to query for data and will rely on the automation of the information system, while other applications are able to influence the processing.

\section{Conclusions}
The presented request driven way of visualization allows scientists to interact with their data in a conceptual way, because all the administration is implicitly taken care of.
The processing and storage is performed in an optimal way and data is shared implicitly.
This frees scientists to focus on what they want to do with the data, instead of how the data is handled and will result in more and faster science.

The current wide field surveys such as KIDS \citep{P157_adassxxi}, VISTA \citep{2007Msngr.127...28A}, the surveys planned for Euclid \citep{2009arXiv0912.0914L} and other astronomical projects such as LOFAR \citep{P009_adassxxi} will produce more data than ever before.
All these projects require the scalability and flexibility provided by query driven visualization and the underlying mechanisms.
Therefore, query driven data visualization is not only a bright possible future, but perhaps an inevitable one.

\acknowledgements
This work was performed as part of the Target project. Target project is supported by Samenwerkingsverband Noord Nederland. It operates under the auspices of Sensor Universe. It is also financially supported by the European fund for Regional Development and the Dutch Ministry of Economic Affairs, Pieken in de Delta, the Province of Groningen and the Province of Drenthe.

\bibliography{P018}

\end{document}